\documentclass[aps,prl,twocolumn,floatfix]{revtex4}
\usepackage{graphicx,units,xspace,amsmath}

\graphicspath{{figures/}}

\newcommand{\micron}{\ensuremath{\unit{\mu m}}\xspace}
\newcommand{\avg}[1]{\ensuremath{\left< #1 \right>}}

\begin{document}

\title{Flux Reversal in a Two-state Symmetric Optical Thermal Ratchet}

\author{Sang-Hyuk Lee}

\author{David G. Grier}

\affiliation{Department of Physics and Center for Soft Matter
  Research\\
New York University, New York, NY 10003}

\date{\today}

\begin{abstract}
  A Brownian particle's random motions can be rectified by a periodic 
  potential energy landscape that alternates between two
  states, even if both states are spatially symmetric.
  If the two states differ only by a discrete translation,
  the direction of the ratchet-driven current can be reversed by
  changing their relative durations.
  We experimentally 
  demonstrate flux reversal in a symmetric two-state ratchet by tracking the
  motions of colloidal spheres moving through large arrays of discrete
  potential energy wells created with dynamic holographic optical tweezers.
  The model's simplicity and high degree of symmetry suggest possible 
  applications in molecular-scale motors.
\end{abstract}

\maketitle

Until fairly recently,
random thermal fluctuations were considered impediments to inducing motion
in systems such as motors.
Fluctuations can be harnessed, however, through mechanisms such as
stochastic resonance \cite{hanggi02} and thermal ratchets \cite{reimann02},
as efficient transducers of input energy into mechanical motion.
Unlike conventional machines, which battle noise, molecular-scale
devices that exploit these processes
actually requite thermal fluctuations to operate.

This article focuses on thermal ratchets in which the random motions of
Brownian particles are rectified by a time-varying potential energy
landscape.
Even when the landscape has no overall slope and thus exerts no
average force, directed motion still can result from 
the accumulation of coordinated impulses.
Most thermal ratchet models break spatiotemporal
symmetry by periodically translating, tilting or otherwise modulating 
a spatially asymmetric landscape \cite{reimann02}.
Inducing a flux is almost inevitable in such systems unless they
satisfy conditions of spatiotemporal symmetry or
supersymmetry \cite{reimann01}.
Even a
spatially symmetric landscape can induce a flux
with appropriate driving
\cite{chen97,kanada99,jones04,lee05}.
Unlike deterministic motors, however, the direction of motion
in these systems can depend 
sensitively
on implementation details.
  
We recently demonstrated a spatially symmetric three-state thermal ratchet
for micrometer-scale colloidal particles implemented with arrays of
holographic optical tweezers, each of which constitutes a discrete
potential energy well \cite{lee05}.
Repeatedly displacing the array first by one third of a lattice 
constant and then by two thirds breaks spatiotemporal
symmetry in a manner that induces a flux.
Somewhat surprisingly, the \emph{direction} 
of motion depends sensitively on the
duration of the states relative to the time required for a particle
to diffuse the inter-trap separation \cite{lee05}.
The induced flux therefore can be canceled or even reversed by
varying the rate of cycling, rather than the direction.
This approach builds upon the pioneering demonstration of
unidirectional flux induced by a spatially asymmetric 
time-averaged optical ratchet \cite{faucheux95,harada04}, and of
reversible transitions driven by stochastic resonance in
a dual-trap rocking ratchet \cite{mccann99,dykman00}.

\begin{figure}[tbh]
  \centering
  \includegraphics[width=.85\columnwidth]{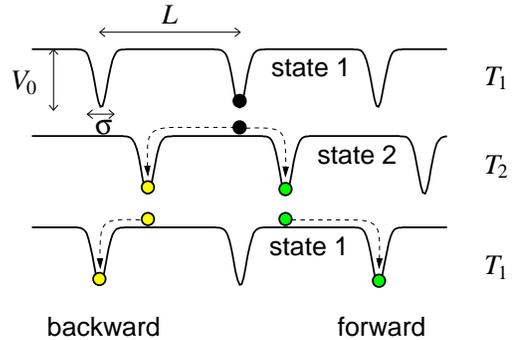}
  \caption{One complete cycle of a 
    spatially-symmetric two-state ratchet potential
    comprised of discrete potential wells.}
  \label{fig:schematic}
\end{figure}
 
Here, we demonstrate flux induction and flux reversal in a
symmetric \emph{two-state} thermal ratchet implemented with dynamic
holographic optical trap arrays \cite{dufresne98,curtis02}.
The transport mechanism for this two-state ratchet is
more subtle than our previous three-state model in that the
direction of motion is not easily intuited from the protocol.
Its capacity for flux reversal in the absence of external loading, 
by contrast, can be inferred immediately
by considerations of spatiotemporal symmetry.  
This also differs from the three-state
ratchet \cite{lee05} and the rocking double-tweezer
\cite{mccann99,dykman00}
in which flux reversal results from a finely tuned balance
of parameters.

Figure~\ref{fig:schematic} schematically depicts how the
two-state ratchet operates.  
Each state consists of a pattern of discrete optical traps,
modeled here as Gaussian wells of width $\sigma$ and depth $V_0$,
uniformly separated by a distance $L \gg \sigma$.
The first array of traps is extinguished after time $T_1$ and replaced 
immediately with a
second array, which is displaced from the first by $L/3$.
The second pattern is extinguished after time $T_2$ and replaced
again by the first, thereby completing one cycle.

If the potential wells in the second state overlap those in the
first, then trapped particles are handed back and forth
between neighboring traps as the states cycle, and no motion results.
This also is qualitatively different from the three-state ratchet, which
deterministically transfers particles forward under comparable conditions,
in a process known as optical peristalsis \cite{koss03,lee05}.
The only way the symmetric two-state ratchet can induce motion is if trapped
particles are released when the states change and then diffuse freely.

\begin{figure}[tbh]
  \centering
  \includegraphics[width=.85\columnwidth]{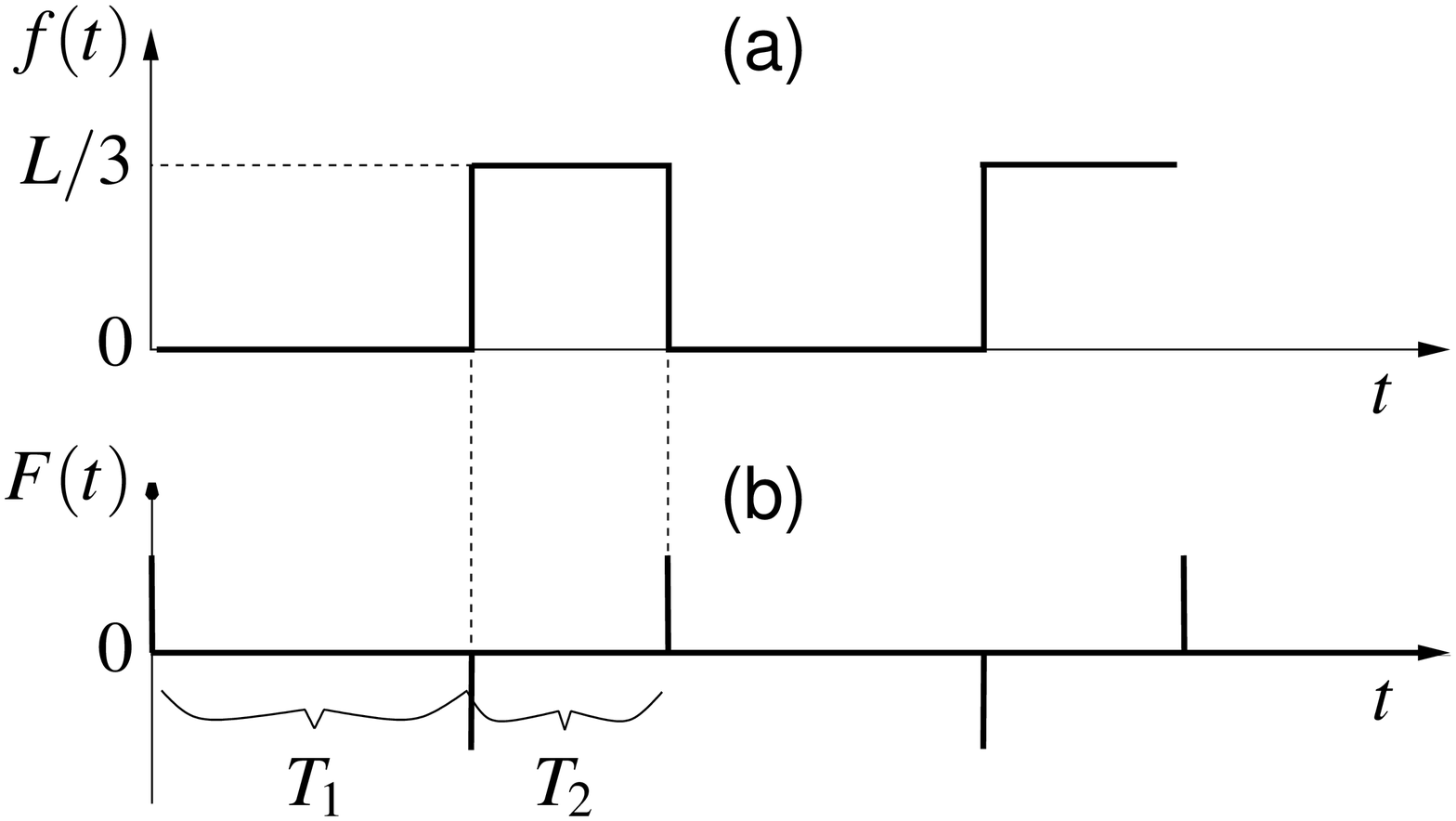}
  \caption{(a) Displacement function $f(t)$.
           (b) Equivalent tilting-ratchet driving force,
           $F(t) = - \eta \dot{f}(t)$.
           }
  \label{fig:f}
\end{figure}

The motion of a Brownian particle in this system can be described
with the one-dimensional Langevin equation
\begin{equation}
  \label{eq:langevin}
  \eta \dot{x}(t) = - V^\prime(x(t) -  f(t)) + \xi(t),
\end{equation}
where $\eta$ is the fluid's dynamic viscosity, $V(x)$ is the
potential energy landscape, 
$V^\prime(x) = \partial V(x) / \partial x$ is its derivative, and
$\xi(t)$ is a delta-correlated stochastic force representing thermal noise.
The potential energy landscape in our system is spatially periodic
with period $L$,
\begin{equation}
  V(x+L) = V(x).
\end{equation} 
The time-varying displacement of the potential
energy in our two-state ratchet is described by a periodic function $f(t)$ 
with period $T = T_1 + T_2$, which is plotted
in Fig.~\ref{fig:f}(a). 

The equations describing this
traveling potential ratchet can be recast into the form
of a tilting ratchet, which ordinarily would be implemented
by applying an oscillatory external force to objects on an otherwise
fixed landscape.
The appropriate coordinate transformation,
$y(t) = x(t) - f(t)$ \cite{reimann02},
yields
\begin{equation}
  \label{eq:tilting}
  \eta \dot{y}(t) = - V^\prime(y(t)) + F(t) + \xi(t),
\end{equation}
where $F(t) = - \eta \dot{f}(t)$ is the effective driving force.
Because $f(t)$ has a vanishing mean,
the average velocity of the original problem is the same as
that of the transformed tilting ratchet  
$\avg{\dot{x}} =  \avg{\dot{y}}$,
where the angle brackets imply both an ensemble average 
and an average over a period $T$.

Reimann has demonstrated \cite{reimann01,reimann02}
that a steady-state flux, $\avg{\dot{y}} \ne 0$,
develops in any tilting ratchet that breaks
both spatiotemporal symmetry,
\begin{equation}
  \label{eq:symmetry}
  V(y)   = V(-y) \text{, and }  -F(t) = F(t + T/2),
\end{equation}
and also spatiotemporal supersymmetry,
\begin{equation}
  \label{eq:supersymmetry}
  -V(y)  = V(y + L/2) \text{, and } -F(t + \Delta t)  = F(-t),
\end{equation}
for any $\Delta t$.
No flux results if either of Eqs.~(\ref{eq:symmetry}) 
or (\ref{eq:supersymmetry})
is satisfied.

The optical trapping potential depicted
in Fig.~\ref{fig:schematic} is symmetric
but not supersymmetric. 
Provided that $F(t)$ violates the symmetry condition in
Eq.~(\ref{eq:symmetry}), the
ratchet must induce directed motion.
Although $F(t)$ is supersymmetric, as can be seen in
Fig.~\ref{fig:f}(b),
it is symmetric only when $T_1 = T_2$.
Consequently, we expect a particle current
for $T_1 \ne T_2$.
The zero crossing at $T_1 = T_2$ furthermore
portends flux reversal on either side of the equality.

\begin{figure}[t!]
  \centering
  \includegraphics[width=\columnwidth]{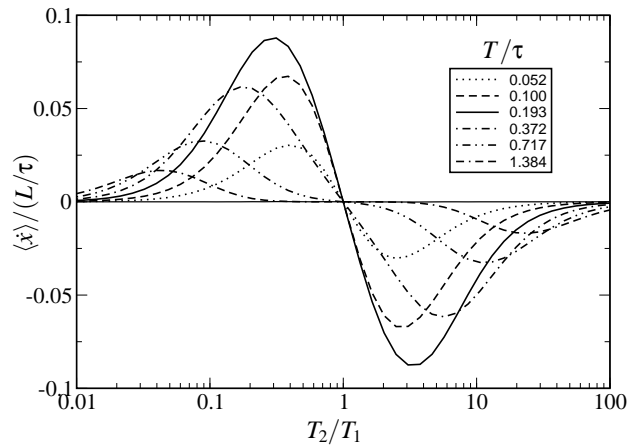}
  \caption{Steady-state drift velocity as a function of the
    relative dwell time, $T_2/T_1$, for 
    $\beta V_0 = 2.75$, $L = 5.2~\micron$,
    $\sigma = 0.65~\micron$, and various values of $T/\tau$.
    Transport is optimized under these conditions by running
    the ratchet at $T/\tau = 0.193$.
    }
  \label{fig:simulation}
\end{figure}

We calculate the steady-state velocity for this system
by solving the master equation associated with Eq.~(\ref{eq:langevin})
\cite{risken89,lee05}. 
The probability for a driven
Brownian particle to drift from position $x_0$ to within $dx$ of position $x$ 
during the interval $t$,
is given by the propagator
\begin{equation}
  \label{eq:propagator}
  P(x, t | x_0, 0) \, dx = e^{\int^{t} L(x,t^\prime) \, dt^\prime} \, \delta(x-x_0) \, dx,
\end{equation}
where the Liouville operator is
\begin{equation}
  L(x,t) = D \, \left(\frac{\partial^2}{\partial x^2} +
    \beta \frac{\partial}{\partial x} V^\prime(x,t) \right),
\end{equation}
and where $\beta^{-1}$ is the thermal energy scale \cite{risken89}.
The steady-state particle distribution $\rho(x)$ is an eigenstate of the
master equation
\begin{equation}
  \label{eq:master}
  \rho(x) = \int P(x, T |x_0,0) \, \rho(x_0) \, dx_0,
\end{equation}
and the associated steady-state flux is \cite{lee05}
\begin{equation}
  \label{eq:flux}
  v = \int \frac{x - x_0}{T} \, \rho(x_0) \, P(x,T | x_0,0) \, dx \, dx_0.
\end{equation}
The natural length scale in this problem is $L$, the inter-trap
spacing in either state.
The natural time scale, $\tau = L^2/(2D)$, is the time required for particles of
diffusion constant $D$ to diffuse this distance.

Figure~\ref{fig:simulation} shows how $v$ varies with $T_1/T_2$ for
various values of $T/\tau$ for experimentally accessible values of $V_0$, $\sigma$, and $L$.
As anticipated, the net drift vanishes for $T_1 = T_2$.
Less obviously, the induced
flux is directed from each well in the longer-duration state
toward the nearest well in the short-lived state.
The flux falls off as $1/T$ in the limit of large $T$ because the
particles spend increasingly much of their time localized in traps.  
It also diminishes
for short $T$ because the particles cannot keep up with the
landscape's evolution.
In between, the range of fluxes
can be tuned with $T$.

\begin{figure}[t!]
  \centering
  \includegraphics[width=0.9\columnwidth]{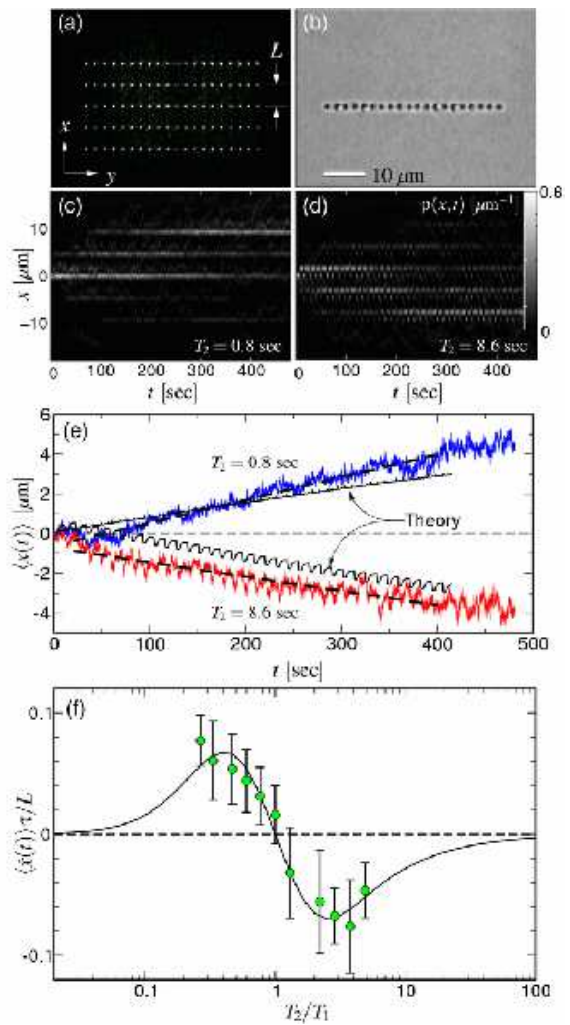}
  \caption{(a) Image of $5 \times 20$ 
    array of holographic optical traps at $L = 5.2 \micron$. 
    (b) Video micrograph of colloidal silica spheres $1.53~\micron$
    in diameter trapped in the middle row of the array at the start
    of an experimental run.
    (c) and (d) Time evolution of the measured probability density for
    finding particles at $T_2 = 0.8~\unit{sec}$ and 
    $T_2 = 8.6~\unit{sec}$, 
    respectively, with $T_1$ fixed at $3~\unit{sec}$. 
    (e) Time evolution of the particles' mean position 
    calculated from the distribution functions in (c) and (d). 
    The slopes of linear fits provide estimates for the induced
    drift velocity, which can be compared with displacements
    calculated with Eq.~(\ref{eq:meanpropagator}) for 
    $\beta V_0 = 2.75$, and $\sigma = 0.65~\micron$.
    (f) Measured drift speed as a function of relative dwell time
    $T_2/T_1$, compared with predictions of Eq.~(\ref{eq:flux}).
    }
  \label{fig:propagator}
\end{figure}

We implemented this model
for a sample of 1.53~\micron diameter colloidal
silica spheres (Bangs Laboratories, lot number 5328)
dispersed in water, using potential energy landscapes
created from arrays of holographic optical traps
\cite{dufresne98,dufresne01a,curtis02,lee05}.
The sample was enclosed in a hermetically sealed glass chamber
roughly 40~\micron thick
created by bonding the edges of a coverslip to a microscope slide,
and was allowed to equilibrate to room temperature ($21 \pm 1^\circ\unit{C}$)
on the stage of a Zeiss S100 2TV Axiovert inverted optical microscope.
A $100\times$ NA 1.4 oil immersion SPlan Apo objective lens was used to
focus the optical tweezer array into the sample and to image the
spheres,
whose motions were captured with
an NEC TI 324A low noise monochrome CCD camera.
The micrograph
in Fig.~\ref{fig:propagator}(a)
shows the focused light from a $5 \times 20$ array of optical traps
formed by a phase hologram projected with a Hamamatsu X7550 
spatial light modulator \cite{igasaki99}.
The tweezers are arranged in twenty-trap manifolds $37~\micron$ long 
separated by $L = 5.2~\micron$.
Each trap is powered by an estimated $2.5 \pm 0.4~\unit{mW}$ 
of laser light at 532~\unit{nm}.  
The particles, which appear in the bright-field micrograph in 
Fig.~\ref{fig:propagator}(b),
are twice as dense as water and sediment to the lower glass surface,
where they diffuse freely in the plane with a measured
diffusion coefficient
of $D = 0.33 \pm 0.03~\unit{\micron^2/sec}$. 
This establishes the characteristic time scale for the system of
$\tau = 39.4~\unit{sec}$, which is quite reasonable for digital video
microscopy studies.
Out-of-plane fluctuations were minimized by focusing the traps at the
spheres' equilibrium height above the wall \cite{behrens03}.

We projected two-state cycles of optical trapping patterns in which the
manifolds in Fig.~\ref{fig:propagator}(a)
were alternately displaced in the spheres' equilibrium plane by $L/3$, 
with the duration of the first state fixed at $T_1 = 3~\unit{sec}$ and 
$T_2$ ranging from 0.8~\unit{sec} to 14.7~\unit{sec}.
To measure the flux induced by this cycling potential energy landscape for one value
of $T_2$,
we first gathered roughly two dozen particles 
in the middle row of traps in state 1, as shown in Fig.~\ref{fig:propagator}(b),
and then projected up to one hundred periods of two-state cycles.
The particles' motions
were recorded as uncompressed digital video streams for
analysis \cite{crocker96}. 
Their time-resolved trajectories then were averaged over the
transverse direction into the probability density, $\rho(x,t) \Delta x$, 
for finding particles within $\Delta x = 0.13~\micron$ 
of position $x$ after time $t$.
We also tracked particles outside the trapping pattern
to monitor their diffusion coefficients and to ensure the absence of
drifts in the supporting fluid.
Starting from this well-controlled initial condition resolves any uncertainties
arising from the evolution of nominally random initial conditions \cite{lee05}.

Figures~\ref{fig:propagator}(c) and (d) show the spatially-resolved 
time evolution of $\rho(x,t)$ for $T_2 = 0.8~\unit{sec} < T_1$ and
$T_2 = 8.6~\unit{sec} > T_1$.
In both cases, the particles spend most of their time localized in traps,
visible here as bright stripes,
occasionally using the shorter-lived traps as springboards to neighboring
wells in the longer-lived state.
The mean particle position $\avg{x(t)} = \int x \, \rho(x,t) \, dx$ advances
as the particles make these jumps, with the associated results plotted in 
Fig.~\ref{fig:propagator}(e).

The speed with which an initially localized state, 
$\rho(x,0) \approx \delta(x)$, advances differs
from the steady-state speed plotted, in Fig.~\ref{fig:simulation}, but still
can be calculated as
the first moment of the propagator,
\begin{equation}
  \label{eq:meanpropagator}
  \avg{x(t)} = \int y \, P(y, t | 0, 0) \, dy.
\end{equation}
Numerical analysis reveals a nearly constant mean speed that agrees
quite closely with the steady-state speed from Eq.~(\ref{eq:flux}).

Fitting traces such as those in Fig.~\ref{fig:propagator}(e) to linear trends
provides estimates for the ratchet-induced flux, which are plotted in
Fig.~\ref{fig:propagator}(f).
The solid curve in Fig.~\ref{fig:propagator}(f) shows excellent agreement with 
predictions of
Eq.~(\ref{eq:meanpropagator}) for $\beta V_0 = 2.75 \pm 0.5$ and
$\sigma = 0.65 \pm 0.05~\micron$.

\begin{figure}[tbh]
  \centering
  \includegraphics[width=0.75\columnwidth]{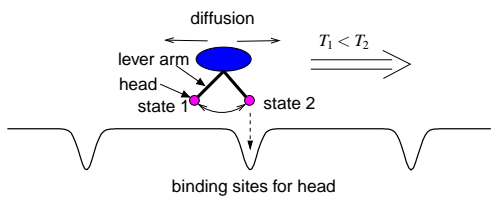}
  \caption{Toy model of diffusive molecular motor.}
  \label{fig:toy}
\end{figure}

Our implementation of the two-state ratchet involves updating the
optical intensity pattern to translate the physical landscape.
However, the same principles can be applied to systems in which
the landscape remains fixed and the object undergoes
cyclic transitions between two states.
Figure~\ref{fig:toy} depicts a model for an active two-state
walker on a fixed physical landscape that is inspired by
the biologically relevant transport of single myosin head groups
along actin filaments \cite{kitamura99}.
The walker consists of a
head group that interacts with localized potential
energy wells
periodically distributed on the landscape.
It also is 
attached to a lever arm that uses an external energy source to
translate the head group by a distance somewhat smaller than
the inter-well separation.
The other end of the lever arm is connected to the payload,
whose viscous drag would provide the leverage necessary to translate
the head group between the extended and retracted states.
Switching between the walker's two states is
equivalent to the two-state translation of the potential
energy landscape in our experiments, and thus would have the
effect of translating the walker in the direction of the shorter-lived
state.
A similar model in which a two-state walker traverses a
spatially asymmetric potential energy landscape yields deterministic
motion at higher efficiency than the present model \cite{julicher97}.
It does not, however, allow for reversibility.
The length of the lever arm and the diffusivity of the motor's body
and payload determine the ratio $T/\tau$ and thus
the motor's efficiency.
The two-state ratchet's direction does not depend on $T/\tau$,
however, even under heavy loading.
This differs from the three-state ratchet
\cite{lee05}, in which $T/\tau$ also controls the 
direction of motion.
This protocol could be used in the
design of mesoscopic motors based on synthetic macromolecules or
microelectromechanical systems (MEMS).

We are grateful for Mark Ofitserov's many technical contributions.
This work was supported by the National Science Foundation through Grant Number
DBI-0233971 and Grant Number DMR-0304906.
S.L. acknowledges support from a Kessler Family Foundation Fellowship


\end{document}